# The Origin of Mutational Epistasis

Jorge A. Vila

IMASL-CONICET, Ejército de Los Andes 950, 5700 San Luis, Argentina.

The interconnected processes of protein folding, mutations, epistasis, and evolution have all been the subject of extensive analysis throughout the years due to their significance for structural and evolutionary biology. The origin (*molecular basis*) of epistasis—the non-additive interactions between mutations—is still, nonetheless, unknown. The existence of a new perspective on protein folding—a problem that needs to be conceived as an "analytic whole"—will enable us to shed light on the origin of mutational epistasis at the simplest level—within proteins—while also uncovering the reasons on *why* the genetic background in which they occur—a key component of molecular evolution—could foster changes in epistasis effects. Additionally, because mutations are the source of epistasis, more research is needed to determine the impact of post-translational modifications—which have the potential to increase the diversity of the proteome by several orders of magnitude—on both mutational-epistasis and protein evolvability. Finally, a protein evolution thermodynamic-based analysis that does not consider specific mutational steps or epistasis effects will be discussed. Our study explores the complex processes behind the evolution of proteins upon mutations, clearing up some previously unresolved issues and providing direction for further research in this area.

**Introduction**

The epistasis phenomenon—which occurs when the combined effects of two or more mutations produce different results from what would be predicted from the addition of their individual effects—has been extensively examined in the literature from many points of view (LiCata & Ackers, 1995; Phillips, 1998; Grant & Grant, 2002; Cordell, 2002; Weinreich *et al*., 2005; Weinreich *et al*., 2006; Ortlund *et al*., 2007; Phillips, 2008; de Visser *et al*, 2011; Breen *et al*, 2012; McCandlish *et al*., 2013; Ashenberg *et al*., 2013; Orgogozo, 2015; Starr & Thornton, 2016; Miton & Tokuriki, 2016; Sailer & Harms, 2017a; Sailer & Harms, 2017b; Adams *et al*, 2019; Domingo *et al*., 2019; Miton *et al*., 2020; Miton *et al*., 2021; Park *et al*., 2022; Jayaraman *et al*, 2022; Buda *et al*., 2023; Diaz *et al*., 2023). Answering this question will not only shed light on basic queries such as *what* the mechanisms that cause mutational epistasis are (Miton & Tokuriki, 2016; Miton *et al*., 2021), but it will also have a profound impact on evolutionary biology since epistasis between mutations plays a critical role on determining evolutionary trajectories (Sailer & Harms, 2017b). The analysis should also allow us to assess the degree to which changes in protein stability and epistasis are independent or weakly correlated phenomena in the evolutionary process (Ashenberg *et al*., 2013). Moreover, whether the epistasis effect could originate from posttranslational modification—a phenomenon that describes an amino acid side-chain modification after their biosynthesis (Mukherjee *et al*., 2007; Chen *et al*., 2010; Khoury *et al*., 2011; Garay *et al*., 2016; Spoel, 2018; Ramazi & Zahiri, 2021; Martin & Vila, 2021; Vila, 2022-2023)—needs to be assessed since, if this were the case, the problem would no longer be limited to the 20 naturally occurring amino acids, and the degree of complexity of the epistasis effect could be greater than anticipated.

For the benefit of the reader, let us start by briefly —but specifically— outlining two fundamental concepts concerning the "protein folding problem" and "protein marginal stability" that have been previously defined.

The "protein folding problem" aims to provide an answer to a *key* question in the structural biology field: *how* does the amino-acid sequence encode its folding? This problem has been under investigation (Vila, 2023a, and references therein) since Anfinsen announced his thermodynamic hypothesis almost 50 years ago (Anfinsen, 1973). Despite the enduring efforts to explain this biological process—essential to understanding both protein function and malfunction—its solution remains unknown (Cramer, 2021; Clementi, 2021; Jones & Thornton, 2022; Vila, 2023c) even when its tridimensional structure can be predicted with high accuracy by state-of-the-art numerical methods such as AlphaFold (Jumper *et al*., 2021). This gives rise to one fundamental conjecture: the protein folding problem should be conceived as an "analytic whole" (Vila, 2023c)—a Leibniz and Kant philosophical notion of space (and time)—, *i.e.*, the one in which "*… Its priority makes it impossible to obtain it by the additive synthesis of previously existing entities*" (Gómez, 1998). From this point of view, methods based mainly on additive pairwise interactions may not give a precise answer to Anfinsen's challenge because such methods consider the 'whole' as *a posteriori* condition rather than a prerequisite (Vila, 2023c). Therefore, the solution demands solving an *n*-body problem (with *n* being the number of amino acids in the sequence). This demand for treating the protein folding as an "analytic whole" is analogous to that needed by numerical/experimental methods aimed at predicting/determining the protein-tridimensional shape, namely, the existence of the structure as a 'whole' as a prerequisite for its resolution (Vila, 2023a).

The "protein marginal stability" hereby refers to the Gibbs free-energy gap between the native state and the first unfolded state (Hormoz, 2013; Vila, 2019; Martin & Vila, 2020; Vila, 2021). In a previous work (Vila, 2019) we have been able, firstly, to demonstrate, based on a statistical-thermodynamics analysis, the existence of an upper limit to the marginal stability of globular proteins, namely ~7.4 Kcal/mol, and secondly, to provide sound arguments that this marginal stability upper bound is valid for any fold-class, sequence or proteins size and, not less important, it is a consequence of a quasi-equilibrium of forces that take place at the minimum of the protein global free energy. Overall, the analysis indicates that: (*i*) the marginal stability of proteins is essentially a consequence of Anfinsen's thermodynamics hypothesis and thus not a consequence of evolution, but rather the physical substrate for evolution to occur, and (*ii*) the marginal stability also seems to be a universal property of biomolecules and macro-molecular complexes (Martin & Vila, 2020). Therefore, this upper-bound limit should not be affected by molecular changes, such as mutations and/or post-translational modifications (Vila, 2021).

Next, we will discuss the *origin* of epistasis at the molecular level—within proteins—as well as the consequences of mutations based on post-translational modification on both epistasis and protein evolution predictability. Subsequently, the question of whether the latter may be accomplished by focusing just on changes in protein marginal stability and by disregarding explicit consideration of the kind of mutations and epistasis effects, and to what extent this may occur, will also be investigated.

**I.- The origin (molecular basis) of epistasis**

As mentioned above, the complex process of mutational epistasis—a phenomenon in which the combined effects of two or more mutations yield outcomes that differ from what would be expected if their individual effects were added—is a ubiquitous phenomenon in the evolution of proteins and, hence, deserves to be deeply analyzed. Despite the broad classes of epistatic interactions described in the literature (Starr & Thornton, 2016) this phenomenon, by definition, perfectly fits in Leibniz and Kant's philosophical notion of space (and time), *i.e.*, one in which the 'whole' is more than the sum of the parts (Gómez, 1998). To put it simply, this means that we need to fully understand *how* the protein amino acid sequence encodes its folding before we can predict the effects of sequence alterations (Vila, 2022, 2023b; 2004). Only then will we be able to comprehend the origin of epistasis at the molecular level, *i.e.*, within proteins. This problem should be added to the many other important issues that structural and evolutionary biology still has to address, such as determining: (*i*) the protein misfolding organization (Serpell *et al.*, 2021); (*ii*) a protein folding pathway (Jones & Thornton, 2022); (*iii*) an accurate estimate of structural and marginal stability changes upon protein point mutations and/or post-translational modifications (Pancotti *et al.*, 2022; Serpell *et al.*, 2021; Buel & Walters, 2022; Vila, 2022; Liu *et al.*, 2024); (*iv*) the protein metamorphosis appearance upon mutations or changes in the milieu (Vila, 2020), *etc*. It is noteworthy to mention that the resolution of each of these problems now rests on resolving a single obstacle: the protein folding problem, or, to put it another way, the solution to an *n*-body problem (Vila, 2023a). Let us now focus on understanding the reason as to *why* the solution to this problem could shed light on the *molecular basis* of the mutational epistasis. As mentioned, if we make a couple of mutations in a protein, we cannot predict the resulting output by the sum of their contributions because of the non-additivity of their effects (epistasis). Is this an unusual phenomenon in nature? Not really. For example, unlike Coulomb forces, van der Waals forces (named after the Dutch physicist Johannes Diderik van der Waals), which are critical for preserving the stability of the native state of the protein (Dill, 1990; Udgaonkar, 2024), are typically not pair-wise additive. *Why*? Simply because the van der Waals forces arising from the interaction between fixed or induced dipoles of molecules are affected by the presence of other molecules nearby (Israelachvili, 1985). As a result, we are unable to get the energy of a molecule net interaction with all other molecules by simply adding *all* its pair potentials. It is quite easy to recognize the definition of mutational epistasis (Miton *et al*., 2020; Jayaraman *et al*., 2022) from this description of the van der Waals forces. In other words, the van der Waals net force and the mutational epistasis must be conceived as an "analytic whole" (Gómez, 1998), *i.e.*, one where the 'whole' is more than the sum of the parts (Vila, 2023c). Before continuing, a word of caution for a proper interpretation of the analogy should be stated. The concept of "nearby" in proteins means 'close' in space, not necessarily in the sequence, although it could also be possible that epistasis may also occur with distant mutations in space (Miton & Tokuriki, 2016). Then we can conclude that mutational epistasis is a phenomenon that is entangled with the protein-folding problem—*how* the amino acids encode their folding—which demands to be solved as an *n*-body problem rather than by pairwise additive interactions, as already discussed in detail (Vila, 2023a).

Once the origin of epistasis has been revealed, it becomes easy to understand high-order epistasis, in which the effect of a mutation is determined by interactions with two or more mutations (Sailer & Harms, 2017b). How is that? Simply because no matter what the "order" of the epistasis is, its origin (molecular basis) is given, as explained above, by *how* the amino acids encode their folding. The idea behind this argument is straightforward: the sequence of amino acids determines the tridimensional structure of a protein in each *milieu* (Anfinsen, 1973). Therefore, a change in the sequence—or the *milieu*—could alter not only the structure and/or function—which usually we refer to as metamorphism (Vila, 2020)—but also the marginal stability of the protein in its native state (Vila, 2022) and thereafter their evolvability (Vila, 2023b). Then, *how* do we determine the impact of a mutation? Simply by finding the lowest accessible Gibbs free-energy conformation at a given *milieu* (the thermodynamic hypothesis), which demands solving protein folding as an *n*-body problem (Vila, 2023a). This conclusion will also enable us to understand more complex scenarios where mutational epistasis arises, *e.g.*, the dependence of the outcome of a mutation on the genetic background (Miton & Tokuriki, 2016). Let us explain the latter. The solution to the protein folding problem—which will provide the molecular basis of epistasis—depends on the *milieu*, a frequently overlooked, although crucial, requirement (Anfinsen, 1973; Vila, 2020). This means there is no reason as to why the mutational epistasis from diverse backgrounds (*milieus*) should be the same. Indeed, this conclusion is supported by evidence indicating that epistasis can be neutral or advantageous in one species while being harmful in another (Breen *et al*., 2012; Miton & Tokuriki, 2016). Thus, alterations to the *milieu*—which are crucial for molecular evolution—may trigger changes in the protein epistasis effect even in the absence of additional mutations. This highlights the importance of considering environmental factors when studying protein-protein interactions and their role in the formation of multiprotein complexes (Jubb *et al*., 2017). Thus, understanding *how* the environment influences protein epistasis can provide valuable insights into the mechanisms driving molecular evolution.

Overall, the determination of the *molecular basis* of mutational epistasis demands the solution of the protein-folding problem in each *milieu*, which is a daunting task. This should encourage researchers to focus their efforts on finding an alternative solution to the protein-folding problem rather than overlooking it; otherwise, progress on *how* mutation affects protein evolvability will be hampered by our incomplete knowledge of the molecular mechanisms of mutational epistasis, and, hence, any strategy for modeling the impacts of mutational epistasis will be rife with ambiguity and imprecise, or the successful solutions will be, at best, severely restricted to specific applications (Weinreich *et al*., 2006; Sailer & Harms, 2017b).

## II.- Mutations, post-translational modifications, and epistasis

The word 'mutation' usually refers to an amino-acid substitution in the protein sequence because of a nucleotide pair replacement within a codon (a nucleotide triplet) (Kimura, 1968), a phenomenon that results from either translational error in protein synthesis or mutational errors in gene replication (Epstein, 1966). To the best of our knowledge, the molecular evolution of proteins has always been studied in terms of 'mutations' in the protein sequence as a result of *only* amino-acid substitutions (Koehl & Levitt, 2002; Bloom *et al*., 2006; Tokuriki *et al*., 2008; Tokuriki & Tawfik, 2009; Socha & Tokuriki, 2013; Otwinowski, 2018; Kurahashi *et al*., 2018; Martin & Vila, 2020; Vila, 2022). Interestingly, post-translational modification (PTM) is a phenomenon that describes an amino acid side-chain modification in some proteins—after their biosynthesis—by a process that goes from the addition of chemical groups (phosphorylation, acetylation, methylation, *etc*.) to polypeptides (ubiquitination) and complex molecules (glycosylation), all of which are reversible processes (Spoel, 2018; Millar *et al*., 2019). As a result, the PTMs have a major impact on the structure, stability, function, and evolvability of proteins, as well as the potential to increase the diversity of the proteome by several orders of magnitude (Shental-Bechor & Levy, 2008; Ellis *et al*., 2012; Chen *et al*., 2010; Garay *et al*., 2016; Spoel, 2018; Millar *et al*., 2019; Ramazi & Zahiri, 2021; Weaver *et al*., 2022). Then, it appears to be a reasonable assumption to think that each post-translational modification (PTM) could give rise to a new amino acid (Vila, 2022). This proposal should not come as a surprise because a change in the side-chain chemistry—which is exactly what a post-translational modification does—is the only thing that distinguishes 19 of the 20 naturally occurring amino acids from one another. From this point of view, the word 'mutation' shall, from here on, refer to a protein sequence change resulting from either an amino-acid substitution—because of a nucleotide pair replacement—or a post-translational modification— because of an amino acid side-chain alteration. In simple terms, any alteration to the side-chain chemistry of any naturally occurring amino acid is referred to as a 'mutation.' As a result, it seems reasonable to think that PTMs could also result in epistasis effects—the non-additive interactions between mutations—in such a way that their origin (molecular basis) should be indistinguishable from those of amino acid substitutions. At this point, it is worth noting that a significant part of the genome is devoted to controlling PTM activation and inactivation (Spoel, 2018; Millar *et al.*, 2019). This phenomenon will pose challenges to the overall understanding of the epistasis effect and, especially to their role in the process of protein evolvability.

We will next look into the potential effects of PTMs on the prediction of protein stability, structure, and evolvability following mutations.

***Impact of the PTMs on****:*

*1.-Protein structure and stability*

As far as we are aware, all existing methods aimed at forecasting the effect of mutations on protein structure and stability (Pancotti *et al.*, 2022; Buel & Walters, 2022; Benevenuta *et al.*, 2023; Kurniawan & Ishida, 2023; Zheng *et al.*, 2024; Liu *et al.*, 2024) are limited to protein sequence substitutions in terms of *only* naturally occurring amino acids. Therefore, all those methods rely on the existence of large data sets—for either training, parameterizing, or assessing protein tridimensional models—that contain an accurate determination of protein-stability changes ($\Delta\Delta G$) after 'amino acid substitutions' (Nikam *et al.*, 2021; Xavier *et al.*, 2021; Tsuboyama *et al.*, 2023). As such, a reevaluation of all these methods will be necessary to forecast the effects of mutations arising from both PTMs and amino acid replacement. Furthermore, the presence of post-translational modification will require an update to the databases, which nowadays solely provide information on amino acid substitutions. The latter is—for the state-of-the-art numerical methods—the greatest weaknesses when it comes to accuracy, and it does not seem that they will be resolved any time soon (Listgarten, 2024).

In contrast to the above, we can confidently assume that the protein marginal stability change upon PTMs (ΔΔG) should always be under the threshold of ~7.4 kcal/mol, beyond which a protein becomes unfolded or nonfunctional (Vila, 2020, 2022). This is because, firstly, PTMs will hardly impact residues at the protein nucleus or nearby, where usually the largest changes in protein destabilization occur (Vila, 2024), and, secondly, the determination of such a threshold was obtained for proteins of any sequence, length, or topology (Vila, 2019; Martin & Vila, 2020; Vila, 2022), and, hence, its value should not be affected by the presence of PTMs.

*2.- The evolution of proteins*

The analysis of the potential impact of post-translational modifications (PTMs) on protein evolvability is an intimidating endeavor because there are endless alternative protein variations, each of which could support a different function, due to the wide range of PTMs and the many residues on which they typically occur (Spoel, 2018). To find an answer to the above-posted challenge, let us start by exploring a simple protein evolution model such as that proposed by Maynard Smith (1970): "…*if evolution by natural selection is to occur, functional proteins must form a continuous network which can be traversed by unit mutational steps without passing through nonfunctional intermediates*…". Implicit in this modeling is that any functional protein that pertains to the protein space obeys Anfinsen's dogma or thermodynamic hypothesis (Anfinsen, 1973). Additionally, every node in this protein-sequence-network model represents a functional protein that differs from its nearest neighbor proteins (nodes) by one mutation brought about by either a PTM or an amino acid substitution. As everyone realizes, following the evolutionary process of a protein or determining the most likely sequence after *n*-mutational steps—even for this plain model of evolution—is a daunting task if epistasis effects are considered. We thus suggested a thermodynamic-based approach (Vila, 2022) as a conceivable way to surmount this formidable challenge. The whole idea behind this approach is to compute the Gibbs free-energy change of the protein marginal stability (ΔG) upon a mutation (ΔΔG) which is verified to be a state function (Vila, 2022). Consequently, the total free-energy change ($\Delta\Delta G_n$), after a certain number of (*n*) mutational steps in the protein sequence space (see Figure 1), will depend *only* on the difference between the protein marginal stability for the wild-type ($\Delta G_{wt}$) and that of the last mutational step ($\Delta G_n$). In other words, $\Delta\Delta G_n = (\Delta G_n - \Delta G_{wt})$ because after *n*-mutations (*n* > 2) $(\Delta G_1 - \Delta G_{wt}) + \sum_{k=2}^{n}(\Delta G_k - \Delta G_{k-1}) = (\Delta G_n - \Delta G_{wt}) = \Delta\Delta G_n$ (Vila, 2022). The *pros* and *cons* of this thermodynamic-based approach follow. The main advantage is that, after *n*-mutational steps, the biophysical properties of any target sequence can be accurately determined—relative to those of the wild-type—without accounting for the epistasis effects along any of the possible $2^n$ evolutionary trajectories (Vila, 2023b). Stated differently, the knowledge of the marginal stability changes of proteins between the initial wild-type sequence and the final target sequence ($\Delta\Delta G_n$) will be enough to accurately determine biophysical parameters for the latter, such as the folding rate or protection factor changes, respectively (Vila, 2024). At this point, it is worth remembering that although thermodynamics itself does not provide us—as statistical mechanics does—with the molecular basis of physical-chemical processes, it can still be applied successfully to analyze complex systems for which molecular analysis is not yet feasible (Hill, 1960), *e.g.*, the protein evolution problem. The major drawback of the chosen thermodynamic-based approach, though, is that it is not possible to determine *why* nature follows one evolutionary trajectory rather than others— which represents an important objective for evolutionary biologists (Weinreich et al., 2006; Sailer & Harms, 2017b). Additionally, this is a problem that becomes aggravated by pieces of evidence showing that most of the pathways leading to a possible sequence might not be accessible (Weinreich *et al*., 2006), making evolution a possibly repeatable process (Miton & Tokuriki, 2016). Yet, it is important to remember that genotypic irreversibility exists (Kaltenbach *et al*., 2015). Despite all of this, the application of the thermodynamic-based approach to identify the most probable protein sequence following *n* mutational steps will be discussed below.

Let us start by assuming that at each mutational site—for a protein of any sequence, length, or topology—there is a total of at least 29 options, *e.g.*, as it results from considering 19 possibilities for amino acid substitution plus 10 out of ~400 existent PTMs; for the latter, we have assumed the existence of only the 10 most frequently seen PTMs, such as phosphorylation, acetylation, ubiquitination, succinylation, methylations, N-linked Glycosylation, O-linked Glycosylation, etc. (Millar *et al*., 2019; Ramazi & Zahiri, 2021). For a protein of 100 residues of any sequence or topology, the above means that the maximum number of possible new sequences ($\Lambda$) will be: $\Lambda \sim 29^{100} \sim 10^{146}$. However, it should be noted that a protein cannot fold faster than ~$10^{-8}$ [sec] $< \tau_0 < $ ~$10^{-5}$ [sec], where $\tau_0$ is the folding speed limit of two-state proteins (Vila, 2023b, and references therein). Consequently, if life on Earth started about a billion (~$10^9$) years ago, this upper bound limit for $\Lambda$ cannot be larger than ~$10^{24}$ (~$10^{21}$). The latter implies that while the enormous number of post-translational modifications adds another layer of complexity to the functional diversity of proteins, it does not increase the accessible upper bound of the protein sequence space ($\Lambda$). A word of caution at this point should be mentioned. To estimate the upper bound limit for $\Lambda$, we have assumed, among other things, that all 20 naturally occurring amino acids could accept PTMs with the same frequency, although it is well-known that this is not the case (Ramazi & Zahiri, 2021). The intricacy of the subject undoubtedly speaks for itself, making any effort to foresee the most likely protein sequence after *n*-mutational steps an extremely challenging task. This is a problem that will be exacerbated if we attempt to explore every possible evolutionary pathway leading to a given target sequence by considering, for example, the epistasis effects.

After all these considerations, we should focus on *how* to identify—from a statistical-thermodynamics analysis—the most probable protein sequence. To do this, we should start by thinking of the Protein Space model (Maynard Smith, 1970) as an ensemble—a mental collection—of all possible functional proteins generated after an arbitrary biological time (see Figure 1). Before doing that, we should keep in mind that the folding rate ($\tau$) is a function of the protein marginal stability ($\Delta G$), *i.e.*, $\tau \approx \exp(\beta \Delta G)$ (Vila, 2023b), and hence the possible number of mutational steps (*n*) that can occur within a given timeframe cannot be predicted with accuracy (Vila, 2024). In spite of this, the most probable protein sequence will be, from a statistical-thermodynamics point of view and without explicit consideration of the epistasis effects, the one from the ensemble of all possible sequences that poses the highest Boltzmann factor, namely, $P_\xi \sim \exp(\beta \Delta\Delta G_\xi)$, where $\xi$ denotes one protein among all possible functional protein sequences generated in a given timeframe (see Figure 1), and $\beta = 1/RT$, with $R$ as the gas constant and $T$ as the temperature in Kelvin degrees. This method enables us to accurately estimate the stability and likelihood of the occurrence of any given protein sequence. Interestingly, the most likely protein sequence—the one with the highest Boltzmann factor—that has been proposed here provides an additional foundation for results from directed evolution studies showing that protein stability promotes evolvability (Bloom et al., 2006; Arnold, 2009). In addition, our result is consistent with the suggestion that "…*evolution of protein function from a given starting point is repeatable*…" (Miton & Tokuriki, 2016). This is a reasonable assumption because the evolution of a given protein sequence must indeed be the most probable one among all possible reachable alternative sequences (see Figure 1). However, the key question is: *what* determines a specific series of mutations in a particular amount of time? There is not certainly a straightforward answer to this inquiry since it will depend on the *milieu*, the specific amino-acid sequence, and the topology of the initial (wild-type) protein. Put differently, the answer lies in understanding *how* a sequence encodes its folding in a particular *milieu* at each mutational step. We attempted to find a workable solution to this challenging issue by approaching it from a thermodynamic point of view, which disregards both the particular series of mutations and the epistasis effects, as opposed to a statistical-mechanics point of view, which considers the molecular interactions in detail, as shown by the analysis of the α-helix-coil transition as a function of pH (Vila, 1986a; 1986b). Although the latter approach provides a more thorough understanding of the underlying mechanisms driving biological processes, both perspectives should be considered as complementary rather than as antagonistic approaches to solving extraordinarily complex problems such as protein evolution. Indeed, it is well-known that any physical-chemical process can be either analyzed from a thermodynamic or statistical mechanics point of view because both will lead to obtaining identical results (Hill, 1969).

Overall, by focusing solely on the result of protein evolution, researchers can gain a broad understanding of how proteins have changed over time. However, this approach may overlook important nuances in the evolutionary process that could provide valuable insights into protein function and adaptation.

**Conclusions**

Understanding the molecular basis of mutational epistasis, which demands determining *how* a sequence encodes its folding by solving an *n*-body problem, will provide foundational knowledge of how mutations interact with each other in a given *milieu* to influence phenotypic outcomes and help clarify how genes change and interact with one another over time. Therefore, changes in the *milieu*, which can be identified by the genetic context in which they take place, and which are essential to molecular evolution, might foster alterations in protein epistasis even in the absence of additional mutations. This highlights the intricate relationship between genetic changes and protein interactions, displaying the dynamic nature of molecular evolution. Understanding these complex interactions can provide important insights into protein evolution and adaptation.

The vast array of posttranslational modifications—estimated to be around 400—adds colossal complexity to the structural and functional diversity of mutated proteins. However, adding posttranslational modification to naturally occurring amino acids as a possible source of mutations would not alter the upper bound of the protein sequence space because proteins fold at a well-limited rate.

To find a reasonable solution to the protein evolution problem, we approached it from a thermodynamic standpoint, which implies ignoring both epistasis effects and specific mutation sequences. Thus, by focusing solely on the result of protein evolution, researchers could gain a broad understanding of how proteins have changed over time. However, this approach will overlook important nuances in the evolutionary process that could provide valuable insights into protein function and adaptation. To surmount this issue, we should perhaps adopt a multidisciplinary approach aimed at finding a comprehensive framework for studying protein evolution. If this approach were feasible, it could enable us to reveal the underlying mechanisms that drive protein evolution (from a statistical-mechanics perspective) and possibly identify new approaches to accurately predict it (from a thermodynamic perspective). Ultimately, by embracing a multidisciplinary approach, we would deepen our knowledge of protein evolution and pave the way for exciting advancements in the field of evolutionary biology.

Even though some of the conclusions derived from our research may or may not be *avant-garde*—given the current state of knowledge regarding protein evolution—we were able to offer an adequate rationale for comprehending certain important challenges, such as the *origin* (molecular basis) of mutational epistasis, which, as far as we know, has lacked a reasonable explanation, as well as ascertained the influence of post-translational modifications on both the protein evolvability and their sequence space size. As a result, we were able to propose strategies to address some of those challenges, intending to improve our understanding and predictability of protein evolution.

**Acknowledgment**

The author acknowledges support from the Institute of Apply Mathematics San Luis (IMASL) and the National Research Council of Argentina (CONICET).

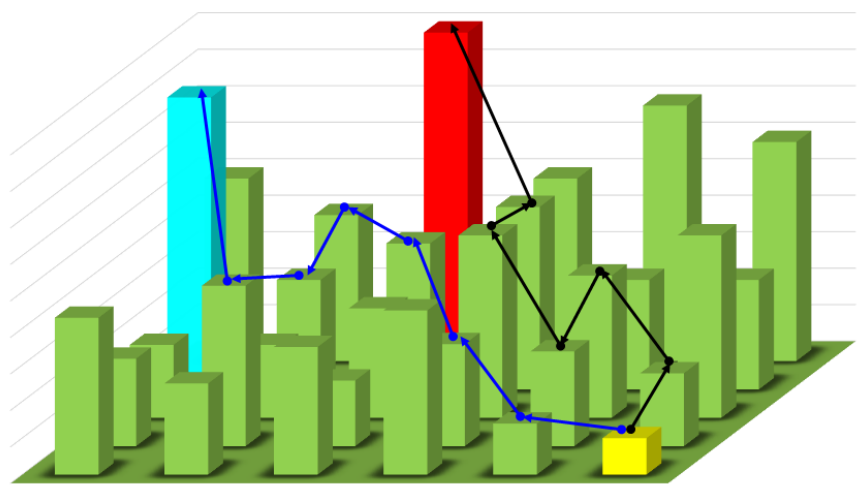

**Figure 1.** A cartoon of the protein sequence space shows two trajectories—black-filled and blue-filled arrows—among all possible ones generated within a given biological timescale. Each bar's height represents the protein sequence's Boltzmann factors attained after a certain number of mutations (see the main text for details) that began with the initial (wild-type) protein, for which the corresponding Boltzmann factor is highlighted as a yellow bar. The bar corresponding to the most probable protein sequence among all shown is highlighted in red. For each of the two indicated final trajectories (red-filled and cyan-filled bars, respectively), there are $2^n$ possible pathways to get there, starting from the wild-type protein sequence. Here, $n$ represents the number of mutational stages, namely, 6 and 7 for the shown trajectories as black and blue-filled arrows, respectively. These pathways represent different combinations of amino acid substitutions, or posttranslational modifications, along the way. The diversity of pathways highlights the complex nature of protein evolution and the potential for multiple ways to achieve the same target sequence. Understanding these pathways could provide valuable insights into the evolutionary processes shaping protein structures and functions.